# Quality Control (QC) of FBK Preproduction 3D Si Sensors for ATLAS HL-LHC Upgrades

D M S Sultan [a,1], Md Arif Abdulla Samy [a,b], J.X. Ye [a,b], M. Boscardin [a,c], F. Ficorella [a,c], S. Ronchin [a,c], and G.-F. Dalla Betta [a,b]

[a] *TIFPA INFN,*
*Via Sommarive 14, 38123 Trento, Italy*

[b] *Department of Industrial Engineering, University of Trento,*
*Via Sommarive 9, 38123 Trento, Italy*

[c] *Fondazione Bruno Kessler,*
*Via Sommarive 18, 38123 Trento, Italy*

E-mail: d.m.s.sultan@cern.ch

ABSTRACT: The challenging demands of the ATLAS High Luminosity (HL-LHC) Upgrade aim for a complete swap of new generation sensors that should cope with the ultimate radiation hardness. FBK has been one of the prime foundries to develop and fabricate such radiation-hard 3D silicon (Si) sensors. These sensors are chosen to be deployed into the innermost layer of the ATLAS Inner Tracker (ITk). Recently, a pre-production batch of 3D Si sensors of 50×50 µm$^2$ pixel geometry, compatible with the full-size ITKPix (RD53B) readout chip, was fabricated. Two wafers holding temporary metal were diced at IZM, Germany, and a systematic QC test campaign was carried out at the University of Trento electronics laboratory. The paper briefly describes the 3D Si sensor design for ATLAS ITk and the required QC characterization setups. It comprises electrical tests (i.e., I-V, C-V, and I-T) of non-irradiated RD53B sensors. In addition, the study of several parametric analyses, i.e., oxide charge density, oxide thickness, inter-pixel resistance, inter-pixel capacitance, etc., are reported with the aid of Process Control Monitor (PCM) structures.

---

[1] Corresponding author.

# Contents

## 1. Introduction

The Large Hadron Collider (LHC) Phase-II upgrade to the higher luminosity ($7.5 \times 10^{34}$ cm$^{-2}$ s$^{-1}$) machine known as the High Luminosity-LHC (HL-LHC) [1] shall go through a complete replacement of current ATLAS detectors. A new generation of radiation-hard silicon sensors [2] shall be loaded in the innermost layer 'L0' of the Inner Tracker (ITk) with different linear, R0, and R0.5 triplet-flavours. Almost 900 3D Si sensors, compatible with the full-size ITKPix (RD53B) readout chip, shall be needed for this purpose's required number of triplets. These sensors should sustain a harsh environment of radiation damage for an integrated luminosity up to 2000 fb$^{-1}$ and a Total Ionizing Dose (TID) of 1 Grad before the replacement. All three foundries (FBK, CNM, and SINTEF) involved in the ATLAS ITk 3D Sensor Collaboration started pre-production fabrication a year ago. Sensors are fabricated on a Si-on-Si wafer, which consists of a high-resistivity float zone active layer of the desired thickness directly bonded to a low-resistivity Czochralski handle wafer. Quality control on FBK pre-production aims to assure the fabrication confidence to ATLAS ITk 3D sensors specifications (AT2-IP-EP-0002). Since no bias structure is allowed on the pixelated 3D detectors, a temporary metal [3,4] is deposited on the readout side of the wafer, shorting all n-type columns to allow for the measurement of the sensor characteristics. Soon after the measurements are done, the metal layer is removed from the foundry side. As a part of quality control, two dedicated wafers holding temporary metal were diced at IZM, Germany, and shipped to the University of Trento (UniTN) electronics laboratory.

A later section of the paper reports a systematic QC electrical characterization of RD53B 3D sensors: Forward I-V, Reverse I-V, C-V, and I-T, along with several parametric studies, i.e., oxide charge density, oxide thickness, inter-pixel resistance, inter-pixel capacitance, etc., that help to understand process parameters, their relationship with device characteristics, and failure mechanisms (whether they exist).

## 2. Sensor Design and Technology

Several processing technologies have been explored for FBK 3D Si sensors [5]. The foreseen HL-LHC extreme radiation damage sets constraints on a small granular pixelated design, with small inter-electrode distances of ~50 µm or less. With such a small pitch design, the dead area due to the 3D electrodes is potentially a concern of the detector efficiency. A small electrode diameter of ~5 µm has been implemented at FBK to address this issue, reducing the ReadOut Chip (ROC) input capacitance. The state-of-art Deep Reactive Ion-Etching (DRIE) process used to create 3D electrodes has an aspect ratio of about 30:1, compatible with active substrate thickness (150 µm). An FBK pre-production batch of 50×50 µm$^2$ geometry was recently released with the temporary metal (Figure 1a). Figure 1b shows the 3D Si sensor cross-schematic fabricated on the 6-inch wafer. The single-sided process has been adopted to assure mechanical integrity, a lower risk of high wafer bow, and bump-bonding complexity.

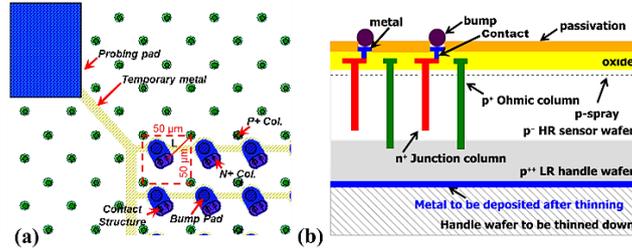

**Figure 1**: (a) micrograph of 50×50 um$^2$ layout geometry. 'L' is the interelectrode distance, ~35 µm, and (b) schematic cross-section (not to scale).

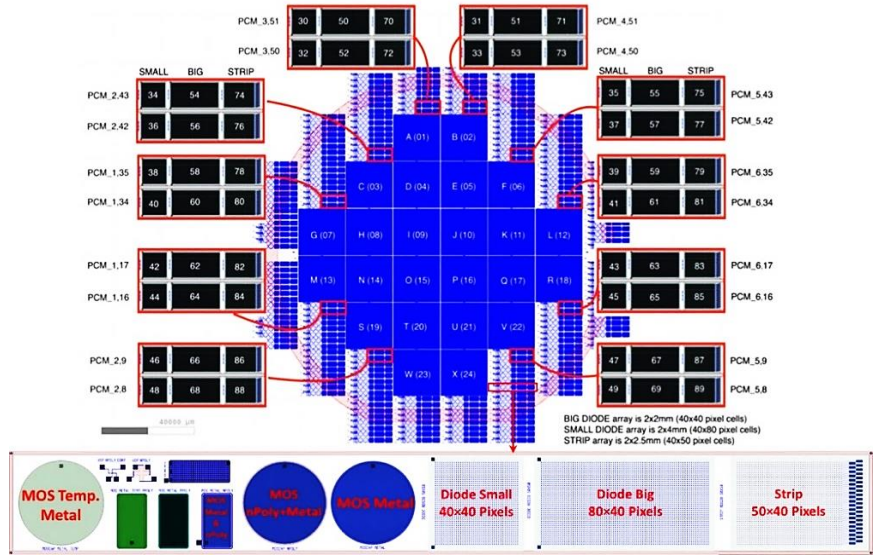

**Figure 2**: Wafer layout details.

As a part of the fabrication steps, DRIE is applied to form P$^+$-columns through 150 µm high-resistive silicon after the necessary oxidation and implantation of p-spray to the sensor front side, penetrating the low-resistive 500 µm handle wafer (thus allowing the sensor back bias). N$^+$-columns are formed at a safety distance of 25 µm from the handle wafer to avoid an early breakdown. Both columns are doped with the respective dopants by the thermal diffusion process, followed by filling (at least partially) with poly-Si. A small extrusion poly-Si cap structure is kept in the column openings for a lower leakage current [6]. Several layers of tetra-ethyl-ortho-silicate (TEOS) oxide are poured into protecting the etched columns. Near N$^+$-columns, metal and n-poly Si are separated by ~150 nm thick thermal oxide and TEOS. Bump pads are placed a little further away to avoid a possible capacitive discharge-driven early breakdown and connected with n-poly Si via the contact structure. Finally, an oxide-nitride passivation layer is deposited over the metal, followed by the openings of bump pads. A temporary metal grid is realized on pixelated sensors and finally removed once the electrical tests are complete. FBK's recent 3D production technology uses stepper lithography, with a minimum feature size of 350 nm and alignment accuracy of 80 nm [7], allowing enough room to fabricate 24 ITkPix-compliant sensors on a 6-inch wafer (see Figure 2). Process Control Monitor (PCM) structures were added around the wafer's periphery, holding different test structures: 3D diode of different dimensions, 3D strips, planar-MOS, etc.

## 3. Experimental Setup

QC setup at UniTN holds a manual probe station without thermal chuck in a dark enclosure. No dry air or $N_2$ flash-inlet is available. A commercial Sensirion temperature & humidity sensor was placed close to the measuring die, and data were logged through an external computer (see Figure 3). The experimental ambient condition remained almost stable (temperature ~21ºC ± 2ºC, and RH <25%). A Keithley 4200 SCS parametric analyzer with four 4210 medium HV-SMUs was used in tests, allowing bias up to ~400V (±210V), with the applied voltage accuracy of ±2 µV and the current sensing resolution of 100 fA. This instrument also has a 4210-CVU unit, allowing C-V measurements of the frequency sweep range between 1kHz-1MHz. This CVU unit can measure 100 fF noise-limited capacitance. A long integration window was applied for measuring small capacitances below 1 nF. For noise-sensitive C-V measurements, a custom measurement window was used: (A/D integration time, 10 PLC) × (Fill Factor, 10).

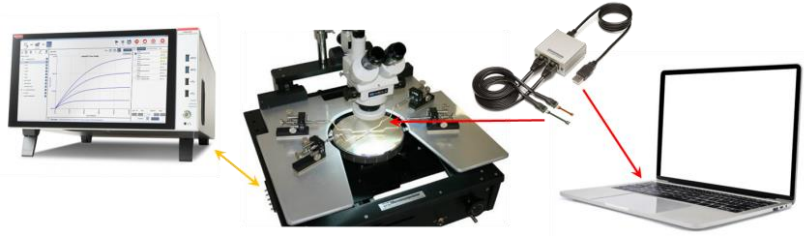

**Figure 3**: UniTN sensor QC probing setup.

An Agilent/Keysight SMU mainframe E5270B with Keysight Precision LCR-Meter E4980A was used to perform electrical tests. HV-SMU of the E5270B mainframe has a precision sensing resolution of 10 fA, and the precision LCR-meter can measure aF capacitance. Wafers were probed in dark conditions with a semiautomatic probe station using a dedicated probe card in the controlled ambiance: temperature ~20ºC±1ºC and RH <5%.

## 4. Experimental Results

This section reports a systematic QC study at UniTN on non-irradiated sensors from wafers W09 and W13 (saw-diced at IZM Germany), holding $N^+$-junction columns shorted by temporary metal grids and routed to contact pads in the periphery. Selective electrical tests were made at FBK on wafers.

### 4.1 Leakage Current

I–V tests were performed at FBK on RD53B sensors and test diodes of different dimensions: small (40×40 cells) and big (40×80 cells), with a current compliance of 100 µA with the applied reverse bias ($V_{bias}$) of 1V step and a delay time of 2 seconds. The following breakdown voltage ($V_{bd}$) definition was used:

$$\frac{I_{leak} @ (V_{bias}+2V)}{I_{leak} @ (V_{bias}+1V)} > 2 \qquad \ldots (1)$$

The approach suggests $V_{bd}$ when there is a 50% leakage rise in a bias step, suitable for a steep avalanche breakdown. A good RD53B sensor reticle is identified with a leakage current <2.5 µA/cm² before breakdown voltage ($V_{bd}$). $V_{bd}$ value has to be 20 V more than the sensor's depletion voltage ($V_{depl}$). Typically, a non-irradiated 3D Si sensor has $V_{depl}$ of a few volts.

UniTN measurements follow a breakdown measurement approach independent of the depletion voltage. They are also suitable for cases where the leakage current shows a smooth and continuous rise due to defects existing in the original wafer or introduced in the silicon lattice after irradiation [8]. Here, $V_{bd}$ is the maximum applied reverse bias where the adimensional function $k(I,V) < k_{bd}$. $k_{bd}$ is a breakdown value that can be retrieved using the following function of a slope ($\frac{\Delta I}{\Delta V}$) of measured I-V data by using Eqn. (2):

$$k(I,V) = \frac{\Delta I}{\Delta V} \cdot \frac{V}{I} \qquad \ldots (2).$$

k<1 represents the sensor ohmic state before the breakdown, and k≫1 corresponds to the actual avalanche characteristics. In this paper, $k_{bd}=4$ is chosen, which also suits a large set of irradiated I-V data of the FBK double-sided 3D sensors [9].

UniTN data were acquired after dicing up to a maximum reverse bias of 150 V or until a hard breakdown was noticed. For example, Figure 4a shows the I-V data for two RD53B sensors of both wafers (W09 and W13), where the leakage current is in good agreement before and after dicing. These results point to a good dicing recipe applied at IZM and minimal contribution of peripheral leakage. The measured leakage before and after dicing is almost two orders of magnitude smaller than the specification. I-V tests were also made for diced diode test structures of different dimensions from wafer W09 (Figure 4b & 4c), showing a hard breakdown greater than 100 V.

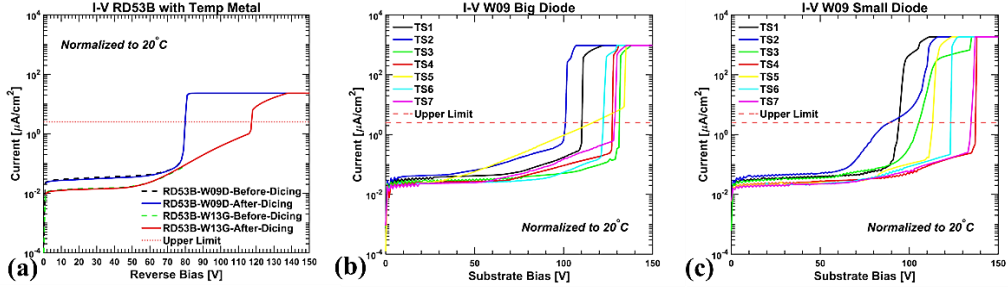

**Figure 4**: I-V curves normalized to 20°C: (a) RD53B sensors of wafer W09 and W13 before and after dicing, (b) W09 big diodes (80×40 pixels) after dicing, and (c) W09 small diodes (40×40 pixels) after dicing.

| DEVICE LEAKAGE CURRENT @ Vrev = 25V | | | |
|---|---|---|---|
| Dev ID | Leakage [A] | VBD [V] | GOOD? |
| A_3,6 | 1.56E-07 | 80.0 | Y |
| B_4,6 | 1.00E-03 | 4.0 | N |
| C_2,5 | 1.88E-07 | 64.0 | Y |
| D_3,5 | 1.75E-07 | 78.3 | Y |
| E_4,5 | 1.00E-03 | 4.0 | N |
| F_5,5 | 1.00E-03 | 4.0 | N |
| G_1,4 | 1.51E-07 | 80.0 | Y |
| H_2,4 | 1.00E-03 | 4.0 | N |
| I_3,4 | 3.25E-08 | 80.0 | Y |
| J_4,4 | 5.19E-08 | 33.0 | Y |
| K_5,4 | 1.00E-03 | 4.0 | N |
| L_6,4 | 1.00E-03 | 5.9 | N |
| M_1,3 | 4.79E-08 | 80.0 | Y |
| N_2,3 | 1.00E-03 | 4.0 | N |
| O_3,3 | 3.20E-08 | 80.0 | Y |
| P_4,3 | 3.15E-08 | 68.0 | Y |
| Q_5,3 | 1.00E-03 | 4.0 | N |
| R_6,3 | 1.00E-03 | 4.0 | N |
| S_2,2 | 3.70E-08 | 80.0 | Y |
| T_3,2 | 3.72E-08 | 80.0 | Y |
| U_4,2 | 1.00E-03 | 4.0 | N |
| V_5,2 | 6.18E-08 | 80.0 | Y |
| W_3,1 | 1.00E-03 | 4.0 | N |
| X_4,1 | 4.01E-08 | 80.0 | Y |

| DEVICE LEAKAGE CURRENT @ Vrev = 25V | | | |
|---|---|---|---|
| Dev ID | Leakage [A] | VBD [V] | GOOD? |
| A_3,6 | 3.02E-04 | 12.0 | N |
| B_4,6 | 1.00E-03 | 4.0 | N |
| C_2,5 | 1.29E-07 | 80.0 | Y |
| D_3,5 | 1.00E-03 | 4.0 | N |
| E_4,5 | 1.65E-07 | 72.0 | Y |
| F_5,5 | 1.67E-07 | 35.1 | Y |
| G_1,4 | 7.65E-08 | 80.0 | Y |
| H_2,4 | 1.00E-03 | 4.0 | N |
| I_3,4 | 1.00E-03 | 4.0 | N |
| J_4,4 | 1.00E-03 | 4.0 | N |
| K_5,4 | 1.65E-07 | 80.0 | Y |
| L_6,4 | 1.43E-07 | 80.0 | Y |
| M_1,3 | 1.00E-03 | 80.0 | N |
| N_2,3 | 7.16E-04 | 21.0 | N |
| O_3,3 | 1.00E-03 | 4.0 | N |
| P_4,3 | 1.00E-03 | 4.0 | N |
| Q_5,3 | 1.10E-07 | 80.0 | Y |
| R_6,3 | 2.17E-07 | 80.0 | Y |
| S_2,2 | 4.11E-08 | 71.3 | Y |
| T_3,2 | 1.07E-07 | 80.0 | Y |
| U_4,2 | 3.31E-08 | 80.0 | Y |
| V_5,2 | 1.00E-03 | 4.0 | N |
| W_3,1 | 6.06E-08 | 80.0 | Y |
| X_4,1 | 6.62E-04 | 4.5 | N |

**Figure 5**: Measured I-V summary of RD53B sensors of different wafers before dicing: (a) W09 and (b) W13.

Figure 5 reports the I-V summary of RD53B sensors from wafers W09 and W13, measured at FBK on the wafer. Yields of wafer W09 and wafer W13 can be estimated as 54% and 50%, respectively. FBK identified an RD53B sensor as good in the case of $V_{bd} > 25$ V and leakage $< 2.5$ µA/cm$^2$ (corresponding to 9.6 µA for the considered sensor area). The bias range was limited to 80 V for a time constraint. Interestingly, many sensors report $V_{bd}$ at 80 V, ascribed from the FBK $V_{bd}$ definition that fails to find smooth breakdowns.

Figure 6a and 6b report the I-V data summary of RD53B sensors, measured at UniTN after dicing. An RD53B sensor is marked good when it has $V_{bd} > 30$ V and leakage $< 2.5$ µA/cm$^2$. The determination of $V_{bd}$ based on k-function allows finding the sensor's actual avalanche-driven breakdown (~50-70 V). Sensors' operation beyond $V_{bd}$ can input a higher noise to ROC and may also lead to sensor surface-quality degradation by employing electrical stress-driven oxide-charge enhancement over time. For wafer W09, the yield appears precisely the same at 54% after dicing, whereas for wafer W13, the yield drops to 38%. It can be anticipated from the saw dicing uncertainties: (1) broken and rough edges that may lead to form micro-cracks and creates peripheral leakage channel into the active sensor volume, and (2) surface damage during handling, a possible source of point-defects. For example, damages were noticed in the W13 RD53B 'F' sensor (Figure 6c), indeed exhibiting an early breakdown.

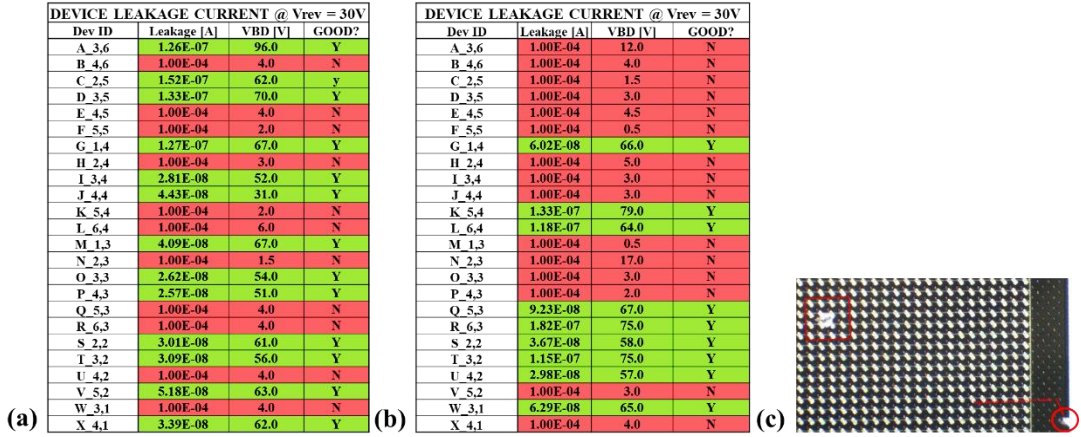

**Figure 6**: Measured I-V summary of RD53B sensors of different wafers after dicing: (a) W09 and (b) W13. (c) An RD53B reticle 'F' micrograph of wafer W13 after dicing.

### 4.2 Leakage Stability

Possible time evolution of leakage current is an important operational aspect of 3D Si sensors at ATLAS ITk. The same sensors in Figure 4a were studied for leakage stability by applying a reverse bias voltage of 30 V for 48 hours at UniTN. The ambient temperature and relative humidity were logged: 20.5 ºC ± 1ºC and ~19.6±3.1%. The leakage fluctuation was estimated by using Eqn. (3):

$$F = \frac{max(I) - min(I)}{average\ (I)} \quad \ldots (3).$$

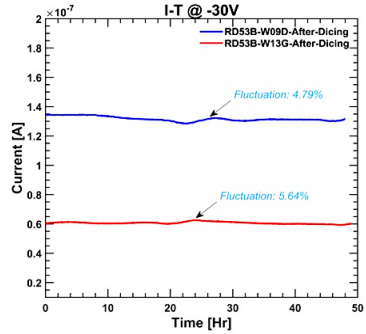

**Figure 7**: Measured RD53B I-T data of different wafers: W09 and W13.

Figure 7 reports the stability test report. The estimated fluctuation of both sensors was ~5%, well below the specification (25%), and can be anticipated from the variation in the ambient conditions.

### 4.3 Bulk Capacitance

Bulk capacitance measurements were performed to define the sensor depletion voltage. During the C-V tests of both RD53B sensors and test diodes, an AC signal was chosen with an amplitude of 100 mV. The frequency was set to 10 kHz. For example, Figure 8a shows RD53B sensors $1/C^2$-V data from different wafers after dicing. The reverse bias was systematically applied with a 1 V increment up to a large reverse bias (till breakdown or dissipation factor <1). The lateral inter-electrode full depletion is reached at a few volts of substrate bias (~2-3 V). The capacitance continues decreasing up to large voltage due to the depletion of the p-spray layer at the surface. A second knee is noticed at ~60 V. A possible explanation is that a weak electric field can be expected underneath the non-passing through $N^+$-column, requiring a sufficiently large bias to deplete the gap region. The depletion summary of all RD53B sensors of both wafers can be found in Figures 8b and 8c. Here, $V_{depl}$ is determined by the interception of two fitted linear segments. The uncertainty of values accounts for the instrument sensing accuracy, errors of linear fits, and the ramp step chosen for substrate biasing. The maximum uncertainty of the reported value is 20%.

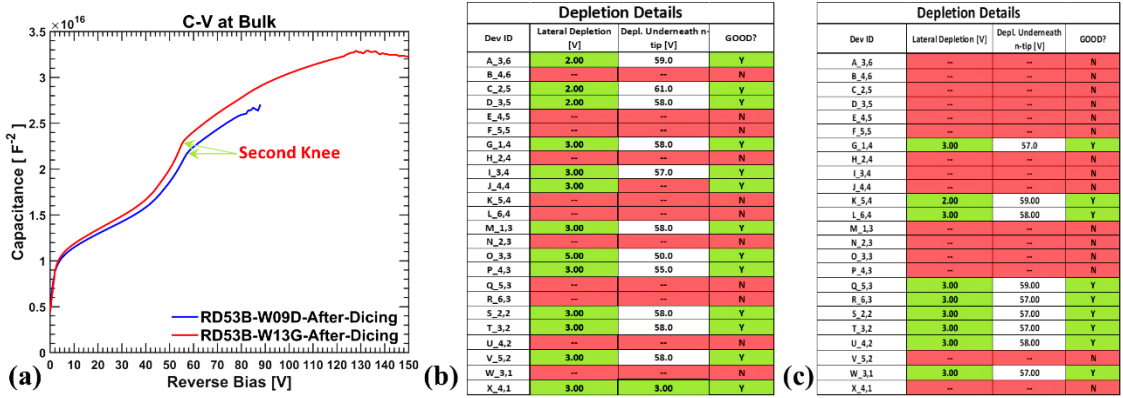

**Figure 8**: (a) Bulk C-V plot of RD53B sensors after dicing, and C-V summary after dicing: (b) W09 and (c) W13.

## 4.4 Forward Current

The forward I-V test is an essential step of sensor QC. It addresses the parasitic series resistance that can play a role in charge collection when it is sufficiently high (> 100 kΩ). To address parasitic series resistance ($R_s$), the measured forward I-V data (up to 1 mA) was fitted with the numerical model of ideal current-voltage relation:

$$I_D = I_S \left[ exp\left(\frac{V_D - (R_S \times I_D)}{\eta \times V_T}\right) - 1 \right] \quad \ldots (4),$$

Rearranging,

$$V_D = \eta \times V_T \times ln\left[\left(\frac{I_D}{I_S} + 1\right) + (R_S \times I_D)\right] \quad \ldots (5).$$

Here, $I_D$ is the diode current for an applied voltage, $V_D$. $I_S$ is the reverse saturation current (typically, $10^{-15}$ A for Si diode), and $V_T$ is the thermal voltage (0.0254 V for 21 ºC). η is the ideality factor that accounts for the effect of electron-hole recombination in the depleted volume. If recombination is negligible, the ideality coefficient η≅1, whereas it can otherwise vary in a range (1 < η ≤ 2).

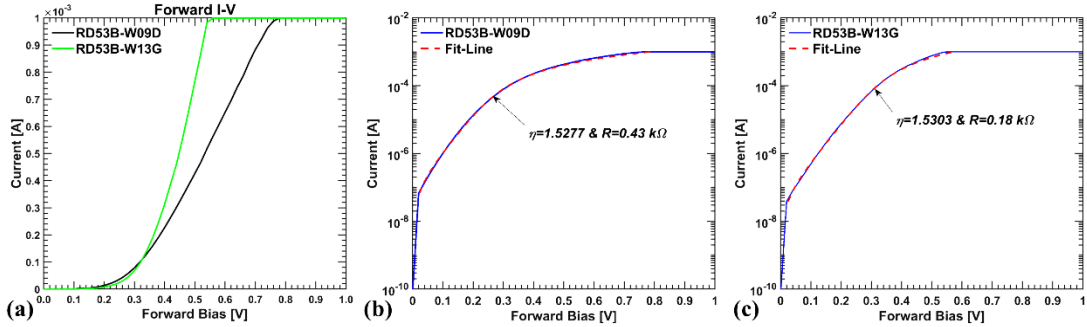

**Figure 9**: (a) Forward I-V plot of RD53B sensors from both wafers. The numerical fit model of the parasitic resistance was applied on: (b) RD53B-W09D & (c) RD53B-W13G.

Figure 9a shows the measured forward I-V data of RD53B sensors. A numerical fitting model was developed in a MATLAB framework for Eqn. 4 & 5 and applied to the measured forward I-V data using a robust fit-option (i.e., Bisquare). The fit lines are also shown in Figures 9b & 9c. The parasitic series resistance has been several 100 Ω only (negligible), while the ideality factor was found to be ~1.5.

## 4.5 Inter-pixel Resistance

Interpixel resistance ($R_{int}$) measurement allows the information of the p-spray isolation implanted between electrodes. A strip test structure was chosen for the test. The central strip was used for sensing current ($I_{int}$). Two neighboring strips were shorted together (Figure 10a), and a systematic voltage sweep ($V_{app}$) was made between -3V to 3V with a step of 0.2V. Substrate reverse bias was applied from 5 V to 60 V with a 5 V step. A long integration window was used. $R_{int}$ is calculated by using Eqn. (6):

$$R_{int} = 2/(dI_{int}/dV_{app}) \qquad \ldots (6).$$

The ambient temperature was ~21.5 °C, and relative humidity was below 20%. The remaining strips of the strip structure were kept floating during the tests. Inter-pixel resistance was normalized from the number of pixels per strip (Figure 10b). As expected, the interpixel resistance is the order of several GΩ (well above the specification) and is independent of the substrate reverse bias.

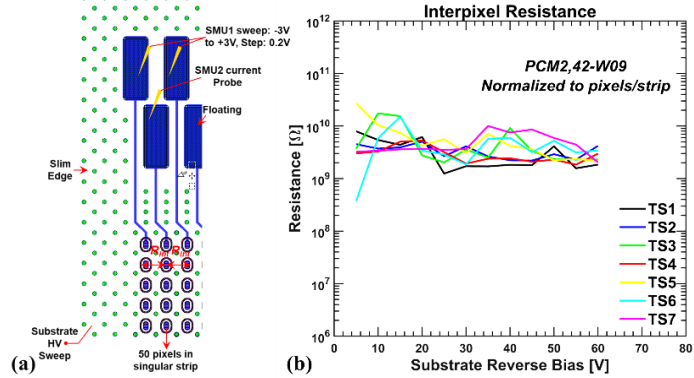

**Figure 10**: (a) Experimental setup, and (b) Inter-pixel resistance of PCM2,42-W09 strip.

### 4.6 Inter-pixel Capacitance

Inter-pixel capacitance ($C_{int}$) measurement quantifies the input capacitance of ROC, which impacts noise. The same strip structure of section 4.5 was chosen for the test. Figure 11a shows the experimental setup schematics. 4210 CVU along HV-SMUs was used to probe the inter-strip capacitance. An input AC signal of 100 mV amplitude and 10 kHz frequency was applied from the central strip to neighboring strips in a parallel mode for a substrate bias (varied from 5 V to 100 V with a 5 V step). The rest neighboring strips were kept floating, which would cause a 10% additional contribution to measured capacitance. The maximum uncertainty would vary by 20% of the reported values, considering the instrumentation errors.

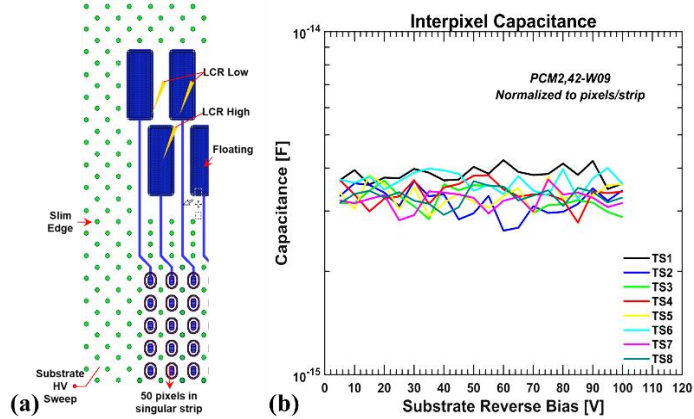

**Figure 11**: (a) Experimental setup, and (b) Inter-pixel capacitance of PCM2,42-W09 strip.

The measured interstrip capacitance has been only a few 100 fF per strip, which makes the measurement very challenging. The test needs precise open corrections to eliminate parasitic capacitance from the cables and the whole setup. Particular attention is required for placing several probe needles onto a tiny area on strip pads that generates large stray capacitances whose fluctuations caused by minimal movements of the probe holders are much higher than the value to be measured. Inter-strip capacitance has been normalized to the number of pixels per strip. Inter-pixel capacitance is ~4 fF (Figure 11b) and is independent of applied substrate bias (as expected).

## 4.7 Surface Parameters

Dedicated planar MOS structures (with nPoly&Metal, Metal, and TempMetalOnly gate electrodes) were studied to quantify the sensor's surface conditions from both wafers. Substrate bias varied from -150 V to 150 V with 1 V step. Figure 12a shows three distinctive regions: accumulation, depletion, and inversion. The depletion state has been spread over a large bias range, ascribed from the continuous p-spray depletion between electrodes. A careful finding of flat-band voltage from the measured C-V data, material work-function values, and the FBK process parameters (i.e., substrate doping, etc.) allows for estimating oxide charge and thickness, summarized in Figure 12b. Values are in excellent agreement with the FBK process. Oxide charge density was at most ~$10^{11}$ cm$^{-2}$, indicating a good surface quality.

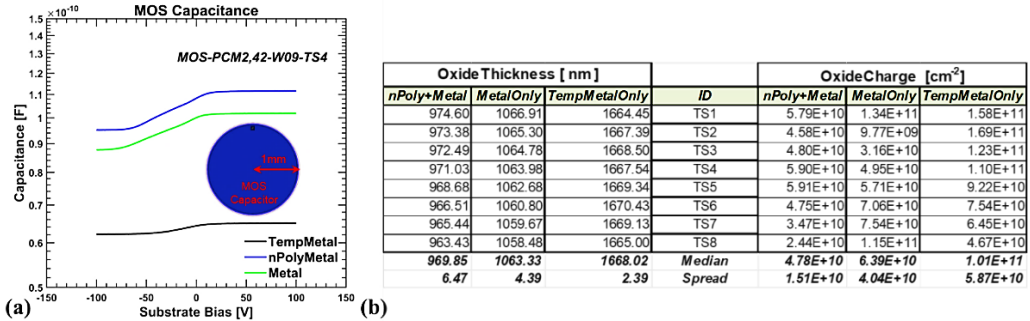

**Figure 12**: (a) C-V plot of PCM2,42-W09 different MOS structures for the different substrate biases. (b) Summary of estimated oxide thickness and oxide charge.

## 5. Conclusions

This paper reports FBK 50×50µm$^2$ 3D pre-production batch QC studies for non-irradiated diced pixel sensors compared with data measured on the wafer at FBK. Leakage current, leakage stability, and depletion voltage are found within the ATLAS ITk sensor specification and are in good agreement before and after dicing. Parasitic series resistance, interpixel resistance, and interpixel capacitance of FBK 3D sensor production are negligible. The lower yield found for wafer W13 is ascribed to dicing problems. Functional studies (i.e., CCE) and investigations on irradiation candidates shall be explored soon.

## Acknowledgments

This work was partially funded by the Italian National Institute for Nuclear Physics (INFN), Projects RD_FASE2 and FASE2_ATLAS (CSN1), and by the H2020 project AIDA-2020, GA No. 654168.